\documentclass[a4paper, amsfonts, amssymb, amsmath, reprint, showkeys, nofootinbib]{revtex4-1}
\usepackage[english]{babel}
\usepackage[utf8]{inputenc}
\usepackage[colorinlistoftodos, color=green!40, prependcaption]{todonotes}
\usepackage{amsthm}
\usepackage{mathtools}
\usepackage{physics}
\usepackage{xcolor}
\usepackage{graphicx}
\usepackage[left=23mm,right=13mm,top=35mm,columnsep=15pt]{geometry} 
\usepackage{adjustbox}
\usepackage{placeins}
\usepackage[T1]{fontenc}
\usepackage{lipsum}
\usepackage{csquotes}
\usepackage[pdftex, pdftitle={Article}, pdfauthor={Author}]{hyperref} 
\usepackage{booktabs}

\begin{document}


\title{Influential Functionals}

\author{M.K. Horton}
    \email[Email address: ]{mkhorton@lbl.gov}
    \affiliation{Materials Science Division, Lawrence Berkeley National Laboratory}

\date{\today}

\begin{abstract}
Learning about density functional approximations (DFAs), or approximations for the exchange-correlation functional, can be intimidating. Density Functional Theory is now one of the primary simulation tools for the practicing chemist or materials scientist, and its accuracy relies upon an appropriate choice of DFA. Over the past decades, there has been extensive research effort to find better DFAs, and there is now a large body of literature to read through for someone learning about DFAs for the first time. In this brief report, I share an analysis that suggests which functionals and publications have been the most influential, as a potential reading list to new scientists in this area. Here, ``influential'' is defined as ``likely to have informed the design of another functional'', and not simply a measure of number of citations, or how much that functional has been used for practical applications. This analysis is not claimed to be complete.
\end{abstract}

\maketitle

\section{Introduction}

This brief report assumes some prior familiarity with Density Functional Theory and density functional approximations, and specifically of exchange-correlation functionals. It is not intended to be a rigorous work, or to be an alternative to a good literature review (of which there are many!), but simply to list which functionals and publications have been highly influential, so as to provide a reading list for those new to the field.

Here, ``influential'' refers to how much a functional has somehow informed the design of new functionals, rather than simply how often it has been cited: the latter is good measure of its overall impact, and how much that functional has been used and has been practically beneficial, but does not necessarily capture the importance of the functional for the development of new functionals. Indeed, many pioneering publications in functional development have not seen many citations, since they have not seen much practical usage but were instead stepping stones to the development of more widely-adopted functionals.

The other motivation for writing up this analysis is to demonstrate a somewhat less biased method of performing a literature review through the combination of routine data retrieval and analysis methods.

\section{Reading List}

The most influential functionals are listed in Tables I, II, III, and IV for those published before 1990, 1990--2000, 2000--2010, and 2010--present, respectively.

In addition, Table V shows papers that are highly cited by these publications, but which do not themselves propose a functional of their own; that is, these are important background reading about the DFT method or key concepts therein.

\section{Method}

The \texttt{libxc}\cite{marques2012libxc,lehtola2018recent} library is a carefully curated, and well-documented, library of exchange-correlation functionals ready for use in other codes. It is difficult to overstate the achievement that is \texttt{libxc} or its value to the community as a resource. For the analysis here, the digital object identifiers (DOIs) of functionals used in \texttt{libxc} are collected, and their references obtained via the Crossref API\cite{crossref}.

A directed graph is constructed with each DOI as a node and a citing relationship as an edge. Link ranking algorithms PageRank\cite{page1999pagerank} and HITS\cite{kleinberg1999authoritative} are applied. All graph analysis is performed using the \texttt{networkx}\cite{hagberg2008exploring} library.

In the tables, ``PR\#'' refers to the overall Pagerank rank index, ``A\#'' refers to the overall ``authorities'' rank index, as defined by the HITS algorithm, and ``Citations'' refers to the total number of citations for that publication.

In total, 398 references are retrieved from the \texttt{libxc} library. The resulting graph, including all references of all 398 publications, has 7366 nodes and 20818 edges. At a paper a day, reading all relevant publications would take over two decades, so this author suggests an effort to rank these papers is well-motivated!

There are some caveats with this data. First, it may not be complete, since \texttt{libxc} may be missing some functionals. Secondly, information on the references of some publications were not available via Crossref, and were ommitted from the analysis. Finally, some references required additional Crossref queries to obtain a DOI, and this process can provide an incorrect DOI in some instances. While this means that some important publications have been ommitted, these caveats are not expected to significantly change the overall ranking of those publications that have been included.

\section{Data Availability Statement}

Data has been uploaded to MPContibs and is available at \url{https://contribs.materialsproject.org/projects/influential_functionals}.

\begin{figure}
    \centering
    \includegraphics[width=3in]{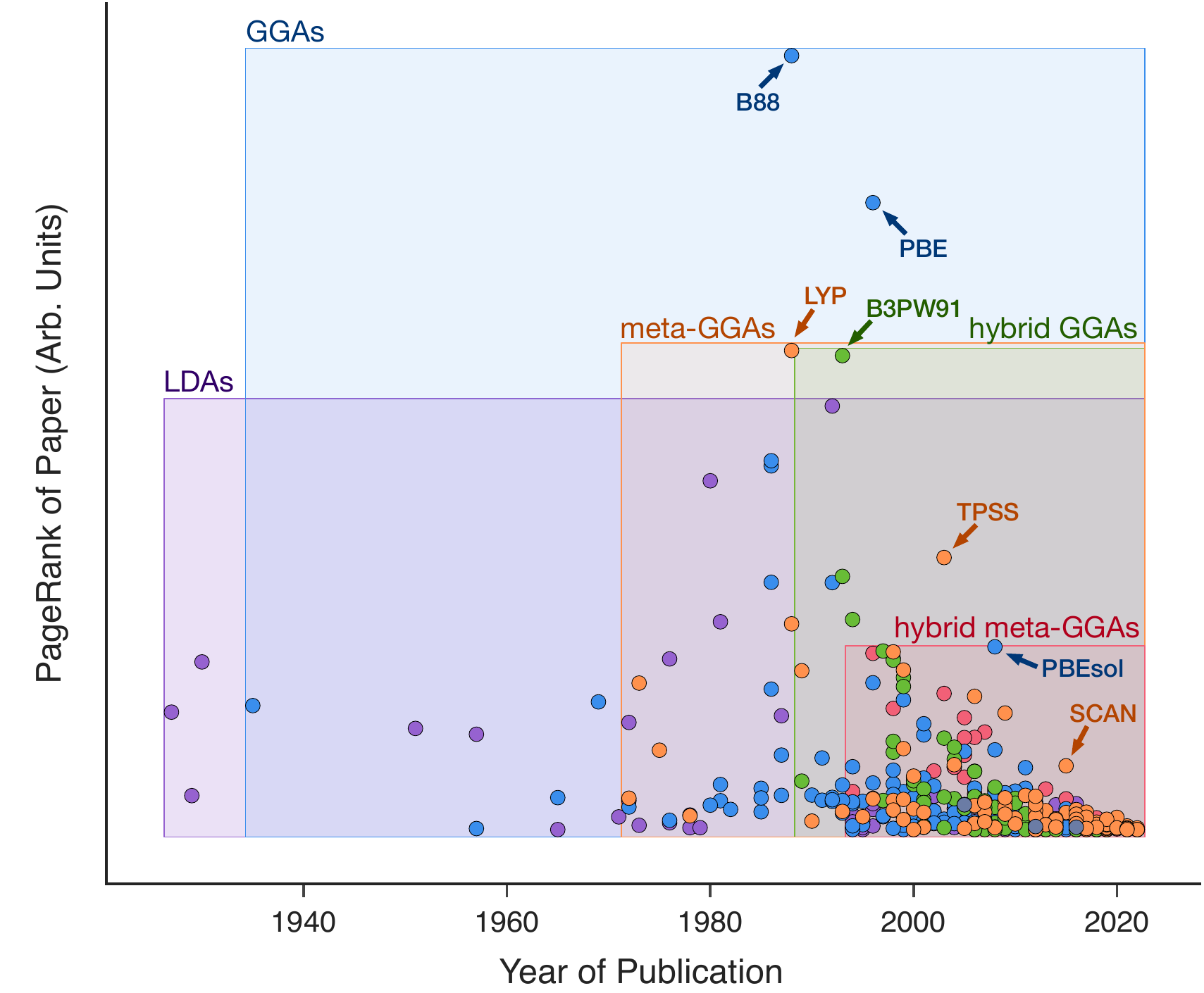}
    \caption{A graph of the PageRank of a given publication (a proxy for its influence), against year of publication. Shaded regions show when specific types of functional were introduced.}
    \label{fig:citations_year}
\end{figure}

\begin{figure}
    \centering
    \includegraphics[width=3in]{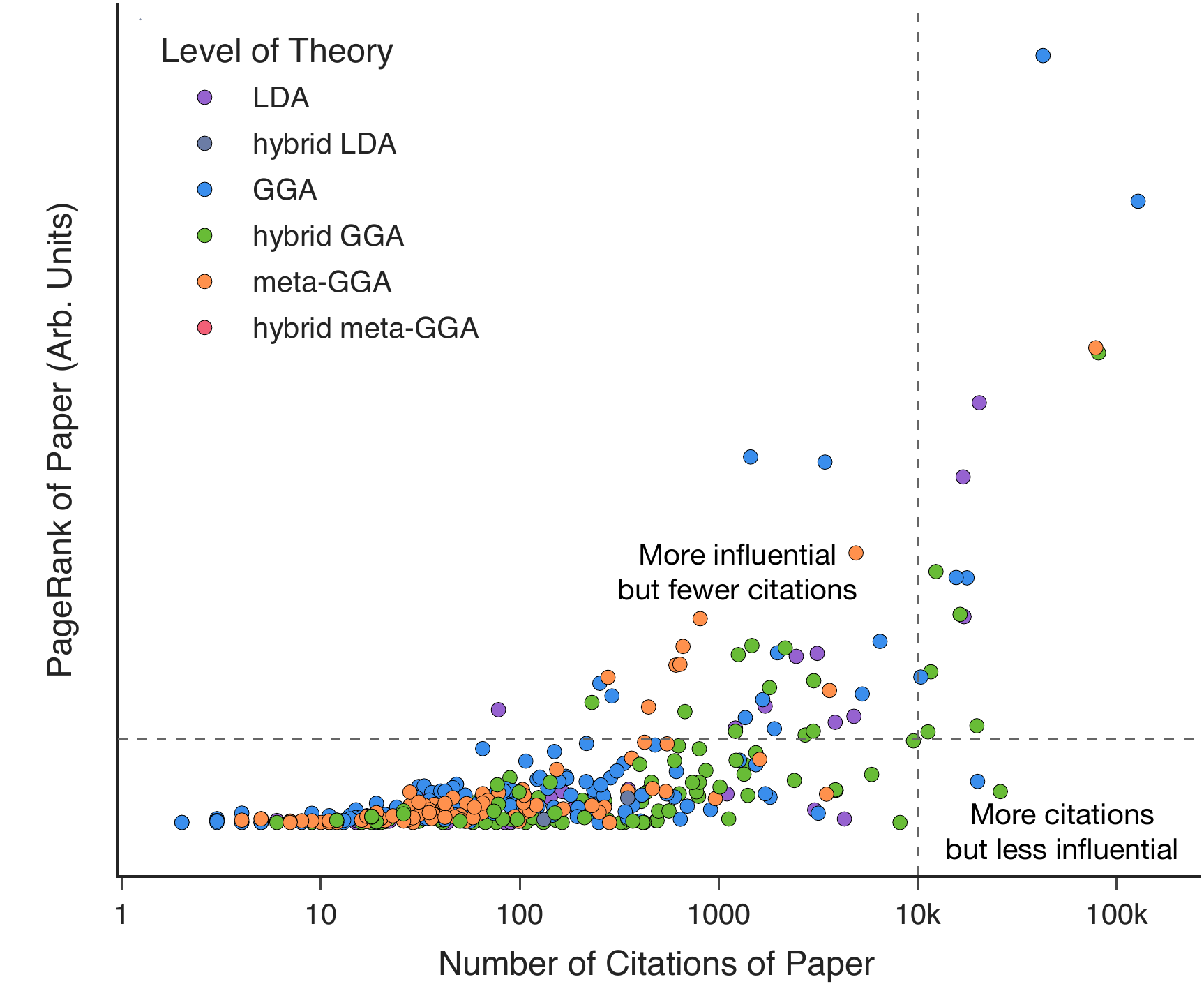}
    \caption{A graph of the PageRank of a given publication (a proxy for its influence) against its overall number of citations.}
    \label{fig:citations_rank}
\end{figure}

\section*{Author's Note}

This brief report originated purely from personal curiosity and was written up for fun. It is only shared since it may be of some legitimate interest to the community, but with the understanding that this is only a quick, preliminary analysis and this report has not been peer-reviewed. I would be grateful to accept corrections for revision. This paper and analysis are shared under CC-BY, so if it's useful for someone to reuse, please do so.

\bibliography{references}

\begin{table*}
\caption{Influential functionals from before 1990.}
\centering
\begin{tabular}{rrrp{1in}lp{3.2in}}
\toprule
PR\# & AR\# & Citations & Label & Theory & Citation \\
\midrule
1 & 2 & 42576 & B88 & GGA & \href{https://doi.org/10.1103/PhysRevA.38.3098}{Becke, A. D. Density-functional exchange-energy approximation with correct asymptotic behavior. Physical Review A 38, 3098–3100 (1988).} \\
4 & 4 & 78367 & LYP, HFLYP, CS & meta-GGA & \href{https://doi.org/10.1103/PhysRevB.37.785}{Lee, C., Yang, W. \& Parr, R. G. Development of the Colle-Salvetti correlation-energy formula into a functional of the electron density. Physical Review B 37, 785–789 (1988).} \\
9 & 53 & 1443 & B86 & GGA & \href{https://doi.org/10.1063/1.450025}{Becke, A. D. Density functional calculations of molecular bond energies. The Journal of Chemical Physics 84, 4524–4529 (1986).} \\
10 & 21 & 3411 & PW86 & GGA & \href{https://doi.org/10.1103/PhysRevB.33.2800}{Perdew, J. P. \& Yue, W. Accurate and simple density functional for the electronic exchange energy: Generalized gradient approximation. Physical Review B 33, 8800–8802 (1986).} \\
11 & 10 & 16880 & VWN & LDA & \href{https://doi.org/10.1139/p80-159}{Vosko, S. H., Wilk, L. \& Nusair, M. Accurate spin-dependent electron liquid correlation energies for local spin density calculations: a critical analysis. Canadian Journal of Physics 58, 1200–1211 (1980).} \\
16 & 37 & 15571 & P86 & GGA & \href{https://doi.org/10.1103/PhysRevB.33.2822}{Perdew, J. P. Density-functional approximation for the correlation energy of the inhomogeneous electron gas. Physical Review B 33, 8822–8824 (1986).} \\
26 & 27 & 17052 & PZ & LDA & \href{https://doi.org/10.1103/PhysRevB.23.2048}{Perdew, J. P. \& Zunger, A. Self-interaction correction to density-functional approximations for many-electron systems. Physical Review B 23, 5048–5079 (1981).} \\
27 & 121 & 805 & B88 & meta-GGA & \href{https://doi.org/10.1063/1.454274}{Becke, A. D. Correlation energy of an inhomogeneous electron gas: A coordinate‐space model. The Journal of Chemical Physics 88, 1053–1062 (1988).} \\
34 & 76 & 3119 & GL & LDA & \href{https://doi.org/10.1103/PhysRevB.13.4274}{Gunnarsson, O. \& Lundqvist, B. I. Exchange and correlation in atoms, molecules, and solids by the spin-density-functional formalism. Physical Review B 13, 4274–4298 (1976).} \\
36 & 89 & 2447 & - & LDA & \href{https://doi.org/10.1017/S0305004100016108}{Dirac, P. A. M. Note on Exchange Phenomena in the Thomas Atom. Mathematical Proceedings of the Cambridge Philosophical Society 26, 376–385 (1930).} \\
39 & 57 & 608 & BR89 & meta-GGA & \href{https://doi.org/10.1103/PhysRevA.39.3761}{Becke, A. D. \& Roussel, M. R. Exchange holes in inhomogeneous systems: A coordinate-space model. Physical Review A 39, 3761–3767 (1989).} \\
45 & 382 & 277 & GEA4 & meta-GGA & \href{https://doi.org/10.1139/p73-189}{Hodges, C. H. Quantum Corrections to the Thomas–Fermi Approximation—The Kirzhnits Method. Canadian Journal of Physics 51, 1428–1437 (1973).} \\
48 & 227 & 252 & B86\_MGC, B86\_R & GGA & \href{https://doi.org/10.1063/1.451353}{Becke, A. D. On the large‐gradient behavior of the density functional exchange energy. The Journal of Chemical Physics 85, 7184–7187 (1986).} \\
55 & 466 & 290 & HERMAN & GGA & \href{https://doi.org/10.1103/PhysRevLett.22.807}{Herman, F., Van Dyke, J. P. \& Ortenburger, I. B. Improved Statistical Exchange Approximation for Inhomogeneous Many-Electron Systems. Physical Review Letters 22, 807–811 (1969).} \\
57 & 133 & 1660 & TFVW, VW & GGA & \href{https://doi.org/10.1007/BF01337700}{Weizs\"acker, C. F. v. Zur Theorie der Kernmassen. Zeitschrift f\"ur Physik 96, 431–458 (1935).} \\
\bottomrule
\end{tabular}
\end{table*}

\begin{table*}
\caption{Influential functionals from between 1990 and 2000.}
\centering
\begin{tabular}{rrrp{1in}lp{3.2in}}
\toprule
PR\# & AR\# & Citations & Label & Theory & Citation \\
\midrule
3 & 1 & 127934 & PBE & GGA & \href{https://doi.org/10.1103/PhysRevLett.77.3865}{Perdew, J. P., Burke, K. \& Ernzerhof, M. Generalized Gradient Approximation Made Simple. Physical Review Letters 77, 3865–3868 (1996).} \\
5 & 5 & 80876 & B3PW91 & hybrid GGA & \href{https://doi.org/10.1063/1.464913}{Becke, A. D. Density‐functional thermochemistry. III. The role of exact exchange. The Journal of Chemical Physics 98, 5648–5652 (1993).} \\
6 & 7 & 20350 & PW & LDA & \href{https://doi.org/10.1103/PhysRevB.45.13244}{Perdew, J. P. \& Wang, Y. Accurate and simple analytic representation of the electron-gas correlation energy. Physical Review B 45, 13244–13249 (1992).} \\
15 & 16 & 12327 & BHANDH, BHANDHLYP & hybrid GGA & \href{https://doi.org/10.1063/1.464304}{Becke, A. D. A new mixing of Hartree–Fock and local density‐functional theories. The Journal of Chemical Physics 98, 1372–1377 (1993).} \\
17 & 36 & 17648 & PW91 & GGA & \href{https://doi.org/10.1103/PhysRevB.46.6671}{Perdew, J. P. et al. Atoms, molecules, solids, and surfaces: Applications of the generalized gradient approximation for exchange and correlation. Physical Review B 46, 6671–6687 (1992).} \\
25 & 8 & 16292 & B3LYP & hybrid GGA & \href{https://doi.org/10.1021/j100096a001}{Stephens, P. J., Devlin, F. J., Chabalowski, C. F. \& Frisch, M. J. Ab Initio Calculation of Vibrational Absorption and Circular Dichroism Spectra Using Density Functional Force Fields. The Journal of Physical Chemistry 98, 11623–11627 (1994).} \\
30 & 22 & 1465 & B97 & hybrid GGA & \href{https://doi.org/10.1063/1.475007}{Becke, A. D. Density-functional thermochemistry. V. Systematic optimization of exchange-correlation functionals. The Journal of Chemical Physics 107, 8554–8560 (1997).} \\
31 & 11 & 660 & GVT4, VSXC & meta-GGA & \href{https://doi.org/10.1063/1.476577}{Van Voorhis, T. \& Scuseria, G. E. A novel form for the exchange-correlation energy functional. The Journal of Chemical Physics 109, 400–410 (1998).} \\
32 & 20 & 2155 & BC95, B86B95, ... & hybrid meta-GGA & \href{https://doi.org/10.1063/1.470829}{Becke, A. D. Density‐functional thermochemistry. IV. A new dynamical correlation functional and implications for exact‐exchange mixing. The Journal of Chemical Physics 104, 1040–1046 (1996).} \\
33 & 42 & 1970 & PBE\_R & GGA & \href{https://doi.org/10.1103/PhysRevLett.80.890}{Zhang, Y. \& Yang, W. Comment on “Generalized Gradient Approximation Made Simple”. Physical Review Letters 80, 890–890 (1998).} \\
35 & 17 & 1251 & HCTH\_A, ... & hybrid GGA & \href{https://doi.org/10.1063/1.477267}{Hamprecht, F. A., Cohen, A. J., Tozer, D. J. \& Handy, N. C. Development and assessment of new exchange-correlation functionals. The Journal of Chemical Physics 109, 6264–6271 (1998).} \\
38 & 28 & 637 & PKZB & meta-GGA & \href{https://doi.org/10.1103/PhysRevLett.82.2544}{Perdew, J. P., Kurth, S., Zupan, A. \& Blaha, P. Accurate Density Functional with Correct Formal Properties: A Step Beyond the Generalized Gradient Approximation. Physical Review Letters 82, 2544–2547 (1999).} \\
40 & 13 & 11608 & PBEH & hybrid GGA & \href{https://doi.org/10.1063/1.478522}{Adamo, C. \& Barone, V. Toward reliable density functional methods without adjustable parameters: The PBE0 model. The Journal of Chemical Physics 110, 6158–6170 (1999).} \\
44 & 70 & 10347 & PBE & GGA & \href{https://doi.org/10.1103/PhysRevLett.78.1396}{Perdew, J. P., Burke, K. \& Ernzerhof, M. Generalized Gradient Approximation Made Simple [Phys. Rev. Lett. 77, 3865 (1996)]. Physical Review Letters 78, 1396–1396 (1997).} \\
46 & 41 & 2996 & PBEH & hybrid GGA & \href{https://doi.org/10.1063/1.478401}{Ernzerhof, M. \& Scuseria, G. E. Assessment of the Perdew–Burke–Ernzerhof exchange-correlation functional. The Journal of Chemical Physics 110, 5029–5036 (1999).} \\
\bottomrule
\end{tabular}
\end{table*}

\begin{table*}
\caption{Influential functionals from between 2000 and 2010.}
\centering
\begin{tabular}{rrrp{1in}lp{3.2in}}
\toprule
PR\# & AR\# & Citations & Label & Theory & Citation \\
\midrule
12 & 6 & 4884 & TPSS & meta-GGA & \href{https://doi.org/10.1103/PhysRevLett.91.146401}{Tao, J., Perdew, J. P., Staroverov, V. N. \& Scuseria, G. E. Climbing the Density Functional Ladder: Nonempirical Meta–Generalized Gradient Approximation Designed for Molecules and Solids. Physical Review Letters 91, (2003).} \\
29 & 12 & 6452 & PBE\_SOL & GGA & \href{https://doi.org/10.1103/PhysRevLett.100.136406}{Perdew, J. P. et al. Restoring the Density-Gradient Expansion for Exchange in Solids and Surfaces. Physical Review Letters 100, (2008).} \\
50 & 14 & 1797 & TPSSH & hybrid meta-GGA & \href{https://doi.org/10.1063/1.1626543}{Staroverov, V. N., Scuseria, G. E., Tao, J. \& Perdew, J. P. Comparative assessment of a new nonempirical density functional: Molecules and hydrogen-bonded complexes. The Journal of Chemical Physics 119, 12129–12137 (2003).} \\
52 & 9 & 3595 & M06\_L & meta-GGA & \href{https://doi.org/10.1063/1.2370993}{Zhao, Y. \& Truhlar, D. G. A new local density functional for main-group thermochemistry, transition metal bonding, thermochemical kinetics, and noncovalent interactions. The Journal of Chemical Physics 125, 194101 (2006).} \\
63 & 26 & 443 & REGTPSS, REVTPSS & meta-GGA & \href{https://doi.org/10.1103/PhysRevLett.103.026403}{Perdew, J. P., Ruzsinszky, A., Csonka, G. I., Constantin, L. A. \& Sun, J. Workhorse Semilocal Density Functional for Condensed Matter Physics and Quantum Chemistry. Physical Review Letters 103, (2009).} \\
68 & 15 & 675 & MPW1KCIS, MPWKCIS1K & hybrid meta-GGA & \href{https://doi.org/10.1021/jp045141s}{Zhao, Y., González-García, N. \& Truhlar, D. G. Benchmark Database of Barrier Heights for Heavy Atom Transfer, Nucleophilic Substitution, Association, and Unimolecular Reactions and Its Use to Test Theoretical Methods. The Journal of Physical Chemistry A 109, 2012–2018 (2005).} \\
72 & 34 & 1357 & ITYH\_OPTX, OPTX & GGA & \href{https://doi.org/10.1080/00268970010018431}{Handy, N. C. \& Cohen, A. J. Left-right correlation energy. Molecular Physics 99, 403–412 (2001).} \\
79 & 29 & 19784 & M06, M06\_2X, ... & hybrid meta-GGA & \href{https://doi.org/10.1007/s00214-007-0310-x}{Zhao, Y. \& Truhlar, D. G. The M06 suite of density functionals for main group thermochemistry, thermochemical kinetics, noncovalent interactions, excited states, and transition elements: two new functionals and systematic testing of four M06-class functionals and 12 other functionals. Theoretical Chemistry Accounts 120, 215–241 (2007).} \\
81 & 61 & 1901 & ITYH, ... & GGA & \href{https://doi.org/10.1063/1.1383587}{Iikura, H., Tsuneda, T., Yanai, T. \& Hirao, K. A long-range correction scheme for generalized-gradient-approximation exchange functionals. The Journal of Chemical Physics 115, 3540–3544 (2001).} \\
84 & 25 & 2981 & M05\_2X & hybrid meta-GGA & \href{https://doi.org/10.1021/ct0502763}{Zhao, Y., Schultz, N. E. \& Truhlar, D. G. Design of Density Functionals by Combining the Method of Constraint Satisfaction with Parametrization for Thermochemistry, Thermochemical Kinetics, and Noncovalent Interactions. Journal of Chemical Theory and Computation 2, 364–382 (2006).} \\
85 & 32 & 1211 & PW6B95, PWB6K & hybrid meta-GGA & \href{https://doi.org/10.1021/jp050536c}{Zhao, Y. \& Truhlar, D. G. Design of Density Functionals That Are Broadly Accurate for Thermochemistry, Thermochemical Kinetics, and Nonbonded Interactions. The Journal of Physical Chemistry A 109, 5656–5667 (2005).} \\
86 & 54 & 11244 & HSE03 & hybrid GGA & \href{https://doi.org/10.1063/1.1564060}{Heyd, J., Scuseria, G. E. \& Ernzerhof, M. Hybrid functionals based on a screened Coulomb potential. The Journal of Chemical Physics 118, 8207–8215 (2003).} \\
95 & 56 & 9496 & CAM\_B3LYP & hybrid GGA & \href{https://doi.org/10.1016/j.cplett.2004.06.011}{Yanai, T., Tew, D. P. \& Handy, N. C. A new hybrid exchange–correlation functional using the Coulomb-attenuating method (CAM-B3LYP). Chemical Physics Letters 393, 51–57 (2004).} \\
99 & 59 & 216 & SOGGA & GGA & \href{https://doi.org/10.1063/1.2912068}{Zhao, Y. \& Truhlar, D. G. Construction of a generalized gradient approximation by restoring the density-gradient expansion and enforcing a tight Lieb–Oxford bound. The Journal of Chemical Physics 128, 184109 (2008).} \\
101 & 63 & 478 & AM05 & GGA & \href{https://doi.org/10.1103/PhysRevB.72.085108}{Armiento, R. \& Mattsson, A. E. Functional designed to include surface effects in self-consistent density functional theory. Physical Review B 72, (2005).} \\
\bottomrule
\end{tabular}
\end{table*}

\begin{table*}
\caption{Influential functionals from after 2010.}
\centering
\begin{tabular}{rrrp{1in}lp{3.2in}}
\toprule
PR\# & AR\# & Citations & Label & Theory & Citation \\
\midrule
130 & 64 & 1604 & SCAN & meta-GGA & \href{https://doi.org/10.1103/PhysRevLett.115.036402}{Sun, J., Ruzsinszky, A. \& Perdew, J. P. Strongly Constrained and Appropriately Normed Semilocal Density Functional. Physical Review Letters 115, (2015).} \\
137 & 86 & 107 & APBE, REVAPBE & GGA & \href{https://doi.org/10.1103/PhysRevLett.106.186406}{Constantin, L. A., Fabiano, E., Laricchia, S. \& Della Sala, F. Semiclassical Neutral Atom as a Reference System in Density Functional Theory. Physical Review Letters 106, (2011).} \\
229 & 107 & 157 & SOGGA11 & GGA & \href{https://doi.org/10.1021/jz200616w}{Peverati, R., Zhao, Y. \& Truhlar, D. G. Generalized Gradient Approximation That Recovers the Second-Order Density-Gradient Expansion with Optimized Across-the-Board Performance. The Journal of Physical Chemistry Letters 2, 1991–1997 (2011).} \\
234 & 101 & 142 & MS1, MS2, MS2\_REV, MS2H & hybrid meta-GGA & \href{https://doi.org/10.1063/1.4789414}{Sun, J. et al. Semilocal and hybrid meta-generalized gradient approximations based on the understanding of the kinetic-energy-density dependence. The Journal of Chemical Physics 138, 044113 (2013).} \\
239 & 88 & 736 & M11 & hybrid meta-GGA & \href{https://doi.org/10.1021/jz201170d}{Peverati, R. \& Truhlar, D. G. Improving the Accuracy of Hybrid Meta-GGA Density Functionals by Range Separation. The Journal of Physical Chemistry Letters 2, 2810–2817 (2011).} \\
256 & 190 & 48 & PBEINT & GGA & \href{https://doi.org/10.1103/PhysRevB.82.113104}{Fabiano, E., Constantin, L. A. \& Della Sala, F. Generalized gradient approximation bridging the rapidly and slowly varying density regimes: A PBE-like functional for hybrid interfaces. Physical Review B 82, (2010).} \\
293 & 97 & 464 & M11\_L & meta-GGA & \href{https://doi.org/10.1021/jz201525m}{Peverati, R. \& Truhlar, D. G. M11-L: A Local Density Functional That Provides Improved Accuracy for Electronic Structure Calculations in Chemistry and Physics. The Journal of Physical Chemistry Letters 3, 117–124 (2011).} \\
303 & 131 & 103 & MS0 & meta-GGA & \href{https://doi.org/10.1063/1.4742312}{Sun, J., Xiao, B. \& Ruzsinszky, A. Communication: Effect of the orbital-overlap dependence in the meta generalized gradient approximation. The Journal of Chemical Physics 137, 051101 (2012).} \\
308 & 112 & 235 & N12 & GGA & \href{https://doi.org/10.1021/ct3002656}{Peverati, R. \& Truhlar, D. G. Exchange–Correlation Functional with Good Accuracy for Both Structural and Energetic Properties while Depending Only on the Density and Its Gradient. Journal of Chemical Theory and Computation 8, 2310–2319 (2012).} \\
336 & 127 & 25964 & PBE38 & hybrid GGA & \href{https://doi.org/10.1063/1.3382344}{Grimme, S., Antony, J., Ehrlich, S. \& Krieg, H. A consistent and accurate ab initio parametrization of density functional dispersion correction (DFT-D) for the 94 elements H-Pu. The Journal of Chemical Physics 132, 154104 (2010).} \\
344 & 146 & 99 & MVS, MVSH & hybrid meta-GGA & \href{https://doi.org/10.1073/pnas.1423145112}{Sun, J., Perdew, J. P. \& Ruzsinszky, A. Semilocal density functional obeying a strongly tightened bound for exchange. Proceedings of the National Academy of Sciences 112, 685–689 (2015).} \\
387 & 164 & 53 & Q2D & GGA & \href{https://doi.org/10.1103/PhysRevLett.108.126402}{Chiodo, L., Constantin, L. A., Fabiano, E. \& Della Sala, F. Nonuniform Scaling Applied to Surface Energies of Transition Metals. Physical Review Letters 108, (2012).} \\
418 & 149 & 790 & VV10, LC\_VV10 & hybrid GGA & \href{https://doi.org/10.1063/1.3521275}{Vydrov, O. A. \& Van Voorhis, T. Nonlocal van der Waals density functional: The simpler the better. The Journal of Chemical Physics 133, 244103 (2010).} \\
428 & 2854 & 32 & CHACHIYO, CHACHIYO\_MOD & LDA & \href{https://doi.org/10.1063/1.4958669}{Chachiyo, T. Communication: Simple and accurate uniform electron gas correlation energy for the full range of densities. The Journal of Chemical Physics 145, 021101 (2016).} \\
436 & 209 & 30 & VMT84\_GE, VMT84\_PBE & GGA & \href{https://doi.org/10.1063/1.3701132}{Vela, A., Pacheco-Kato, J. C., Gázquez, J. L., del Campo, J. M. \& Trickey, S. B. Improved constraint satisfaction in a simple generalized gradient approximation exchange functional. The Journal of Chemical Physics 136, 144115 (2012).} \\
\bottomrule
\end{tabular}
\end{table*}

\begin{table*}
\caption{Influential publications without associated functionals in \texttt{libxc}.}
\centering
\begin{tabular}{rrrp{5in}}
\toprule
PR\# & AR\# & Citations & Citation \\
\midrule
2 & 3 & 45996 & \href{https://doi.org/10.1103/PhysRev.140.A1133}{Kohn, W. \& Sham, L. J. Self-Consistent Equations Including Exchange and Correlation Effects. Physical Review 140, A1133–A1138 (1965).} \\
7 & 104 & 5227 & \href{https://doi.org/10.1016/S0092-640X(74)80016-1}{Clementi, E. \& Roetti, C. Roothaan-Hartree-Fock atomic wavefunctions. Atomic Data and Nuclear Data Tables 14, 177–478 (1974).} \\
8 & 19 & 38547 & \href{https://doi.org/10.1103/PhysRev.136.B864}{Hohenberg, P. \& Kohn, W. Inhomogeneous Electron Gas. Physical Review 136, B864–B871 (1964).} \\
13 & 33 & 3077 & \href{https://doi.org/10.1063/1.460205}{Curtiss, L. A., Raghavachari, K., Trucks, G. W. \& Pople, J. A. Gaussian‐2 theory for molecular energies of first‐ and second‐row compounds. The Journal of Chemical Physics 94, 7221–7230 (1991).} \\
14 & 58 & 780 & \href{https://doi.org/10.1103/PhysRevA.32.2010}{Levy, M. \& Perdew, J. P. Hellmann-Feynman, virial, and scaling requisites for the exact universal density functionals. Shape of the correlation potential and diamagnetic susceptibility for atoms. Physical Review A 32, 2010–2021 (1985).} \\
18 & 67 & 937 & \href{https://doi.org/10.1103/PhysRevB.28.1809}{Langreth, D. C. \& Mehl, M. J. Beyond the local-density approximation in calculations of ground-state electronic properties. Physical Review B 28, 1809–1834 (1983).} \\
19 & 18 & 1768 & \href{https://doi.org/10.1063/1.473182}{Curtiss, L. A., Raghavachari, K., Redfern, P. C. \& Pople, J. A. Assessment of Gaussian-2 and density functional theories for the computation of enthalpies of formation. The Journal of Chemical Physics 106, 1063–1079 (1997).} \\
20 & 179 & 402 & \href{https://doi.org/10.1103/PhysRevLett.55.1665}{Perdew, J. P. Accurate Density Functional for the Energy: Real-Space Cutoff of the Gradient Expansion for the Exchange Hole. Physical Review Letters 55, 1665–1668 (1985).} \\
21 & 40 & 122 & \href{https://doi.org/10.1093/oso/9780195092769.001.0001}{Parr, R. G. \& Weitao, Y. Density-Functional Theory of Atoms and Molecules. (1995) doi:10.1093/oso/9780195092769.001.0001.} \\
22 & 105 & 181 & \href{https://doi.org/10.1002/qua.560230605}{Becke, A. D. Hartree-Fock exchange energy of an inhomogeneous electron gas. International Journal of Quantum Chemistry 23, 1915–1922 (1983).} \\
23 & 69 & 12053 & \href{https://doi.org/10.1103/PhysRevLett.45.566}{Ceperley, D. M. \& Alder, B. J. Ground State of the Electron Gas by a Stochastic Method. Physical Review Letters 45, 566–569 (1980).} \\
24 & 24 & 382 & \href{https://doi.org/10.1002/qua.560190306}{Lieb, E. H. \& Oxford, S. Improved lower bound on the indirect Coulomb energy. International Journal of Quantum Chemistry 19, 427–439 (1981).} \\
28 & 23 & 463 & \href{https://doi.org/10.1103/PhysRevA.47.3649}{Chakravorty, S. J., Gwaltney, S. R., Davidson, E. R., Parpia, F. A. \& p Fischer, C. F. Ground-state correlation energies for atomic ions with 3 to 18 electrons. Physical Review A 47, 3649–3670 (1993).} \\
37 & 110 & 1283 & \href{https://doi.org/10.1063/1.456415}{Pople, J. A., Head‐Gordon, M., Fox, D. J., Raghavachari, K. \& Curtiss, L. A. Gaussian‐1 theory: A general procedure for prediction of molecular energies. The Journal of Chemical Physics 90, 5622–5629 (1989).} \\
41 & 111 & 175 & \href{https://doi.org/10.1103/PhysRevA.20.397}{Oliver, G. L. \& Perdew, J. P. Spin-density gradient expansion for the kinetic energy. Physical Review A 20, 397–403 (1979).} \\
42 & 483 & 2080 & \href{https://doi.org/10.1063/1.462066}{Becke, A. D. Density‐functional thermochemistry. I. The effect of the exchange‐only gradient correction. The Journal of Chemical Physics 96, 2155–2160 (1992).} \\
43 & 39 & 4057 & \href{https://doi.org/10.1063/1.472933}{Perdew, J. P., Ernzerhof, M. \& Burke, K. Rationale for mixing exact exchange with density functional approximations. The Journal of Chemical Physics 105, 9982–9985 (1996).} \\
47 & 30 & 267 & \href{https://doi.org/10.1021/jp035287b}{Lynch, B. J. \& Truhlar, D. G. Small Representative Benchmarks for Thermochemical Calculations. The Journal of Physical Chemistry A 107, 8996–8999 (2003).} \\
49 & 6647 & 2267 & \href{https://doi.org/10.1007/978-1-4757-5714-9}{Mahan, G. D. Many-Particle Physics. (2000) doi:10.1007/978-1-4757-5714-9.} \\
51 & 800 & 31 & \href{https://doi.org/10.1007/978-1-4684-1890-3_36}{Sham, L. J. Approximations of the Exchange and Correlation Potentials. Computational Methods in Band Theory 458–468 (1971) doi:10.1007/978-1-4684-1890-3\_36.} \\
\bottomrule
\end{tabular}
\end{table*}

\end{document}